
\input phyzzx.tex


\def\d{\delta}

\def\t{\theta}

\def\r{\rho}

\def\k{\kappa}
\def\l{\lambda}

\def\s{\sigma}

\def\pa{\partial}
\def\na{\nabla}
\def\hg{\hat g}

\def\ov{\overline}
%
\def\cmp#1{{\it Comm. Math. Phys.} {\bf #1}}

\def\pl#1{{\it Phys. Lett.} {\bf #1B}}

\def\prd#1{{\it Phys. Rev.} {\bf D#1}}

\def\np#1{{\it Nucl. Phys.} {\bf B#1}}

\def\mpl#1{{\it Mod. Phys. Lett.} {\bf A#1}}

%
\REF\call{C.G. Callan, S.B. Giddings, J.A. Harvey, and A. Strominger,
\prd{45} (1992) R1005.}
\REF\hawk{S. W. Hawking,
\cmp{43} (1975) 199.}
\REF\lenny{J.G. Russo, L. Susskind, L. Thorlacius, Stanford preprint (1992)
SU-ITP-92-4, and L. Susskind and L. Thorlacius, Stanford preprint (1992)
SU-ITP-92-12. }
\REF\banks{T. Banks, A. Dabholkar, M.R. Douglas, and M O'Loughlin,  Rutgers
preprint RU-91-54.}
\REF\andy{A. Strominger ITP Santa Barbara preprint (1992)
UCSBTH-92-18,\hfill\break hepth@xxx/9205028.}
\REF\sda{S. P. de Alwis, Colorado preprint, (1992) COLO-HEP-280,\hfill\break
 hepth@xxx/9205069.}
 \REF\bilal{A. Bilal and C. Callan, Princeton preprint (1992)
PUPT-1320,\hfill\break hepth@xxx/9205089.}
\REF\dkd{F. David, \mpl{3} (1988) 1651, J.Distler and H. Kawai, \np{321} (1989)
509.}
\REF\cham{A. Chamseddine,  \pl{256} (1991) 379, and \pl{258} (1991) 97,
 \np{368} (1992) 98; A Chamseddine and Th. Burwick,
preprint hepth 9204002. S. P. de Alwis and J. Lykken, \pl{269} (1991) 264,
J. Russo and A. Tseytlin, SU-ITP-2, DAMTP-1-1992.}

\pubnum {COLO-HEP-284\cr hepth@xxx/9206020}
\date={June, 1992}
\titlepage
\vglue .2in
\centerline{\bf Black Hole Physics from Liouville Theory}
\author{ S.P. de Alwis\foot{dealwis@gopika.colorado.edu}}
\address{Dept. of Physics, Box 390,\break
University of Colorado,\break Boulder, CO
80309}
\vglue .2in
\centerline{\caps ABSTRACT}

In a previous paper it was shown that the quantum consistency conditions for
the dilaton-gravity theory of Callan et al., imply that the cosmological
constant term undergoes a dilaton dependent renormalization, in such a manner
that the theory can be written  as a  Liouville-like theory. In this paper we
discuss the physical interpretation of the solutions of this theory. In
particular we demonstrate explicitly  how quantum corrections tame the  black
hole singularity.  Also under the assumption that in asymptotically Minkowski
coordinates,  there are no incoming or outgoing ghosts, we show that the
Hawking radiation rate  is independent of the number of matter fields and is
determined by the ghost conformal anomaly.

\endpage

In   reference [\call ] a theory of dilaton gravity  which had black hole
solutions resulting from  a collapsing matter shock wave was presented by
Callan, Giddings, Harvey, and Strominger [CGHS]. By adding the
conformal anomaly term to this classical action these authors were able to
discuss Hawking radiation [\hawk] from this black hole. However in the weak
coupling approximation ($e^{2\phi}<<1$ where $\phi$ is the dilaton) in which
these authors worked it was not possible to discuss many interesting questions.
     For instance  it was argued in [\call ] that (when $N$ the number of
matter fields
is very large, so that semi-classical arguments are applicable) the black hole
radiates away all its energy leaving a zero  mass residue behind.  However it
was pointed out [\lenny, \banks] that this theory has a singularity
which prevents the passage from the linear dilaton region (where
$e^{-2\phi}>N>>1$) to the Liouville region (where $N>e^{-2\phi}>>1$). Since the
smooth transition between these two regions was a necessary condition for the
argument of [\call ],
 it was clear that the latter was not valid. Subsequently it was realized
[\andy, \sda, \bilal ] that this singularity is absent when $N<24$. However in
that case semi-classical methods were no longer applicable. One needed the
exact solution of the field equations corrected by the quantum anomaly terms.

In the previous paper by the author [\sda ], and in a paper by Bilal and Callan
[\bilal ], this problem was solved by showing that the quantum consistency
conditions on the CGHS theory require that it  be of the Liouville type. This
happens because the (dilaton dependent) cosmological constant  undergoes a
(dilaton dependent) renormalization. As mentioned in [\sda ] and discussed in
detail  [\bilal ] the new (classical) field equations are  solvable.
The new field space coordinates are explicit (though non-trivial) functions of
the conformal
metric factor and the dilaton, and these solutions therefore are valid to all
orders in the dilatonic coupling $e^{2\phi}$. Of course these are not the full
quantum equations since the effects of  dilaton and graviton loops are not
included.
\foot {Actually since the consistent quantum theory is of the Liouville type it
is possible that all quantum effects in this theory  can be discussed. However
this is a subject for future discussion.} Nevertheless it is sufficient for our
purposes in that we can discuss the effects of Hawking radiation and back
reaction without assuming weak coupling (i.e. being inconsistent).

Let us first summarize  the results of [\sda ]. The classical CGHS
action\foot{$-\l^2$ should be replaced by $\l^2$ in equations (1) and (10) of
reference [5].} is

$$S={1\over 4\pi}\int d^2\s\sqrt{-g}[e^{-2\phi}(R+4(\na\phi )^2+4\l^2)-{1\over
2}\sum_{i=1}^N (\na f^i)^2 ].\eqn\cghs$$

The corresponding quantum field theory is defined by

$$Z=\int {[dg]_g[d\phi ]_g[df]_g\over [Vol.~Diff.]} e^{iS[g,\phi,f]}.\eqn\qft$$

By generalizing an argument first given in [\dkd ],\foot {This generalization
has been considered before by several authors  [\cham ].} this theory can be
written in the gauge fixed form (in the conformal gauge $g=e^{2\r}\hg$)

 $$Z=\int [dX^{\mu}]_{\hg}[df]_{\hg}([db][dc])_{\hg}e^{iI(X,\hg)+iS(f,\hg )+
 iS(b,c,\hg )},\eqn\part$$
  where
  $$I[X,\hg ]=-{1\over 4\pi}\int \sqrt{-\hg}[{1\over 2}
  \hg^{ab}G_{\mu\nu}\pa_aX^{\mu}\pa_bX^{\nu}+\hat R\Phi (X)
+T(X)].\eqn\sigmod$$
Note that all the measures in \part\ are defined with respect to the $2d$
metric $\hg$.
Since the original theory is independent of the fiducial metric $\hg$ the gauge
fixed theory must satisfy the constraints
$$<T_{\pm\pm}+t_{\pm\pm}>=0,\eqn\cons$$
and
$$<T_{+-}+t_{+-}>=0,\eqn\conf$$
where $T_{\mu\nu}$ is the stress tensor for the dilaton-gravity and matter
sectors, and $t_{\mu\nu}$ is the stress tensor for the ghost sector.
In addition  one must have the integrability condition for these constraints,
namely the Virasoro algebra with zero central charge. This means  that $G,\Phi$
and $T$ must obey certain beta-function equations. By solving these equations
under the boundary conditions that the theory reduces to the CGHS theory in the
weak coupling limit, it was shown that the theory is of the form, (choosing
$\hg$ to be the Minkowski metric $\hg_{+-}=-{1\over 2},~\hg_{\pm\pm}=0$)
$$Z=\int [dX][dY][df][db][dc]e^{iS[ X,Y,f]+iS_{ghost}},$$
 where,

 $$S={1\over 4\pi}\int d^2\s[-\pa_{+}X\pa_{-}
X+\pa_{+}Y\pa_{-}Y+\sum_i\pa_{+}f^i\pa_{-}f^i+2\l^2e^{-\sqrt{2\over\k}(X-Y)}].
\eqn\newaction$$
In the above $\k={24-N\over 6}$, and the signs are appropriate for the case
where $\k >0$.\foot{There should be an overall minus sign in front of the
integral sign and $T$ should be multiplied by a factor $-{1\over 2}$ in
equation
(27) of [\sda ].}
The new fields $X$ and $Y$ are defined in terms of the conformal factor and the
dilaton by the following equations.
$$\eqalign{ X=&2\sqrt{2\over\k}\int e^{-2\phi}(1+\k e^{2\phi })^{1\over 2}\cr
=&2\sqrt{2\over\k}[-{1\over 2}e^{-2\phi}(1+\k e^{2\phi})^{1\over 2}-{\k\over
4}\log 2 +{\k\over 2}\phi +\cr &+\log (1+{\k\over 2}e^{2\phi}+(1+\k
e^{2\phi})^{1\over 2})\cr \simeq & -\sqrt{2\over\k}e^{-2\phi}+\sqrt{2\k}\phi
+const.+O(\k e^{2\phi}),\cr}\eqn\xcdt$$
 (the last relation is valid only in the weak coupling limit $e^{-2\phi}>>1$)
$$ Y=\sqrt{2\k }(\r-\k^{-1}e^{-2\phi}).\eqn\ycdt$$

The equations of motion coming from \newaction\ are
$$\pa_+\pa_-f=0,\eqn\feqn$$
$$\eqalign{\pa_+\pa_-X=&\l^2\sqrt{2\over\k}e^{-\sqrt{2\over k}(X-Y)}\cr
           \pa_+\pa_-Y=&\l^2\sqrt{2\over\k}e^{-\sqrt{2\over
k}(X-Y)}.\cr}\eqn\xy$$

Equation \feqn\  gives
$$f=f_+(\s^+)+f_-(\s^-).$$

Subtracting the second from the first in \xy\ we have, $\pa_+\pa_-(X-Y)=0$,
giving
$$X=Y+\sqrt{\k\over 2}(g_+(\s^+)+g_-(\s^-)).$$
Substituting this result into (say) the $Y$ equation we get,

$$\pa_+\pa_-Y=\l^2\sqrt{2\over\k}e^{-(g_+(\s^+ )+g_-(\s^- ))}.$$

 This may be trivially integrated giving the solution,

$$\eqalign{Y=&-\sqrt{2\over\k}(u_+(\s^+)+u_-(\s^-))+\l^2\sqrt{2\over\k}
\int^{\s^+}d\s^+e^{-g_+(\s^+)}\int^{\s^-}d\s^-e^{-g_-(\s^-)}\cr
 &=X-\sqrt{\k\over 2}(g_++g_-).\cr}\eqn\soln$$
 In the above $g_{\pm},u_{\pm}$ are arbitrary (chiral) functions of $\s^{\pm}$.
 $g_{\pm}$ may be set to zero by a conformal transformation and $u_{\pm}$
  have to be determined by the boundary conditions.

These solutions are of exactly the same form as the CGHS solutions to the
classical theory without conformal anomalies. Our observation amounts to the
statement that all the complications due to the latter are hidden in the
relations \xy\ between the fields $X,Y$ and the original fields $\phi
,\r$.\foot{The coordinate choice for which $g_{\pm }$ is zero is the
Kruskal-Szekeres type
one denoted by $x^{\pm}$ in [\call ]. We will however continue to call these
coordinates $\s^{\pm}$.} With this choice we have
$$X=Y=-\sqrt{2\over\k}(u-\l^2\s^+\s^-).\eqn\ssoln$$
where $u=u_++u_-$.
In the semi-classical limit ($e^{-2\phi}>>1$) using the relations \xcdt, and
\ycdt,
we have from the above solution (in this limit $X=Y$ implies  $\phi =\r$),

$$e^{-2\phi}=e^{-2\r}=u-\l^2\s^+\s^-,\eqn\cghssoln$$

which is just the CGHS solution.

The classical metric and curvature are  singular on the curve defined by the
vanishing of the left hand side of \cghssoln. However this singularity is in
the strong coupling limit $e^{\phi}>>1$, where  we have from \xcdt,
$$\eqalign{X=&2\sqrt 2\int e^{-\phi}\left (1+{e^{-2\phi}\over\k}\right
)^{1\over 2}\cr =&2\sqrt 2\left (-e^{-\phi}-{e^{-3\phi}\over 6\k}
+O({e^{-5\phi}\over\k})\right ).\cr}\eqn\strong$$
 Using our solution \ssoln\ and \ycdt\ we then get,
 $$\eqalign{e^{-\phi}\simeq&{1\over2\sqrt\k}(u-\l^2\s^+\s^-),\cr
 \r\simeq&-{1\over\k}(u-\l^2\s^+\s^-).\cr}\eqn\strongcoup$$
Thus the quantum corrected metric $e^{2\r}$ is not singular at the strong
coupling point. The quantum effects have eliminated the classical black hole
singularity.
Let us now work out some of the consequences of our exact solution. By taking
the derivative of \ssoln\ with respect to $\s^{\pm}$ and using \xcdt, and
\ycdt, we have,

$$\eqalign{2e^{-2\phi}(1+\k e^{2\phi})^{1\over 2}\pa_{\pm}\phi
=&-\pa_{\pm}u{\pm}+\l^2\s^{\mp}\cr
=&\k\pa_{\pm}\r+2e^{-2\phi}\pa_{\pm}\phi ,\cr}$$

giving us

$$2e^{-2\phi}\pa_{\pm}\phi=-{(\pa_{\pm}u_{\pm}-\l^2\s^{\pm})\over (1+\k
e^{2\phi})^{1\over 2}},\eqn\dphi$$

$$\k\pa_{\pm}\r=-{(1+\k e^{2\phi})^{1\over 2}-1\over (1+\k e^{2\phi})^{1\over
2}}(\pa_{\pm}u_{\pm}-\l^2\s^{\mp}).\eqn\drho$$

The first equation gives us the trajectory of the apparent horizon ($\pa_+\phi
=0$) of the putative black hole, introduced by Russo, Susskind and Thorlacius
[\lenny ]; i.e.
$$\s^-={1\over\l^2}\pa_+u_+(\s^+).\eqn\hor$$

We can also calculate the curvature.

$$\eqalign{R=8e^{-2\r}\pa_+\pa_-\r =&-{8\l^2(1-(1+\k e^{2\phi})^{1\over
2})\over\k (1+\k e^{2\phi})^{1\over 2}}e^{-2\r}\cr+&4e^{4\phi
-2\r}{(\pa_-u_--\l^2\s^+)(\pa_+u_+-\l^2\s^-)
\over (1+\k e^{2\phi})^2},\cr}\eqn\curv$$

where $e^{2\phi}$ and $e^{-2\r}$ are determined implicitly by \xcdt, \ycdt, and
\ssoln. This expression shows clearly how the curvature singularity of the
classical theory (i.e. the strong coupling singularity at
 $e^{2\phi}\rightarrow\infty$), is tamed by the quantum anomaly corrections;
      specifically by the $O(\k e^{2\phi})$ terms in the denominators. Of
course for $\k <0$ we would have the singularity discovered in
 [\lenny, \banks ].

In order to determine the function $u$ we need to use the constraints \cons.
(\conf\ is not a new equation since it is equivalent to the $\r$ equation of
motion). Calculating the stress tensor for the non-ghost part of the action
from \newaction\ we get,

 $$\eqalign{T_{\pm\pm}=&-{1\over
2}(\pa_{\pm}X\pa_{\pm}X-\pa_{\pm}Y\pa_{\pm}Y)-\sqrt{\k\over 2}\pa^2_{\pm}Y
 +{1\over 2}\sum_i\pa_{\pm}f^i\pa_{\pm}f^i\cr
 =&e^{-2\phi}(4\pa_{\pm}\phi\pa_{\pm}\r-2\pa^2_{\pm}\phi)+{1\over
2}\sum_i\pa_{\pm}f^i\pa_{\pm}f^i+\k(\pa_{\pm}\r\pa_{\pm}\r-\pa^2_{\pm}\r
).\cr}\eqn\stress$$

In the coordinate system that we have chosen $X=Y$, so that substituting the
solution \ssoln\ we get
$$\eqalign{T_{\pm\pm}=&{1\over 2}\sum_i\pa_{\pm}f^i\pa_{\pm}f^i-
\sqrt{\k\over2}\pa_{\pm}Y\cr=&{1\over
2}\sum_i\pa_{\pm}f^i\pa_{\pm}f^i+\pa_{\pm}^2u_{\pm}.\cr}$$
Hence the constraint equations \cons\ become,

$$\pa_{\pm}^2u_{\pm}+{1\over 2}\sum_i\pa_{\pm}f^i\pa_{\pm}f^i+t_{\pm\pm}=0.$$

Following [\call ] we assume that matter falls in in the form of a shock wave
with
${1\over 2}\sum_i\pa_{+}f^i\pa_{+}f^i=a\d(\s^+-\s_0^+),~f_-=0$. Solving the
above equations we then have,

$$\eqalign{u_+=&a_++b_+\s^+-a(\s^+-\s^+_0)\t (\s^+-\s^+_-)-\int\int
t_{++}(\s^+),\cr
u_-=&a_-+b_-\s^--\int\int t_{--}(\s^-).\cr}\eqn\ueqn$$

Now under a conformal transformation
$\s^{\pm}\rightarrow\hat\s^{\pm}=f^{\pm}(\s^{\pm})$,

$$\eqalign{T'_{\pm\pm}(\hat\s )=&\left ({\pa f^{\pm}\over\pa\s^{\pm}}\right
)^{-2}[T_{\pm\pm}(\s )-{\k+N+2\over 2}Df^{\pm}],\cr
t_{\pm\pm}'(\hat\s )=&\left ({\pa f^{\pm}\over\pa\s^{\pm}}\right
)^{-2}[t_{\pm\pm}(\s )-{-26\over 12}Df^{\pm}],\cr}\eqn\trans$$

where $Df$ is the Schwartz derivative  defined by,

$$Df={f'''\over f'}-{3\over 2}\left ({f''\over f'}\right )^2.$$

 The sum of the two stress tensors  is zero in every frame but one cannot put
one or the other to zero except in a particular frame. We will argue that
 the ghost stress tensor should be put to zero in the asymptotically Minkowski
frame which covers all of  space-time.\foot{This argument was first
made in \bilal. However the result for Hawking radiation in that paper is
somewhat different from
ours.} Such a frame is defined  by the transformation (for
$\s^+>\s^+_0$)
 $$\hat\s^+={1\over\l}\log (\l\s^+),~\hat\s^-=-{1\over\l}\log (-\l\s^-).$$

Calculating the Schwartz derivative for this we find $Df^+={1\over 2\s^{+2}},~
Df^-={1\over 2\s^{-2}}$. Plugging this into \trans\ and putting $t'(\hat\s
)=0$ we get,

$$t_{++}=-{26\over 24}{1\over\s^{+2}},~t_{--}=-{26\over 24}{1\over \s^{-2}}.$$
Substituting these values into \ueqn\ we find,

$$\eqalign{u_+=&a_++b_+\s+-a(\s^+-\s^+_0)\t (\s^+-\s^+_0) -{26\over 24}\log
\s^+,\cr u_-=&a_-+b_-\s^--{26\over 24}\log (\s^-).\cr}\eqn\usoln$$
 To get the classical solution of [\call ] with infalling matter we should put
$a_{\pm}=0=b_{\pm}$.
Then we can find the trajectory of the apparent horizon from \hor.
$$\s_H^-={1\over\l^2}(-a\t (\s^+-\s^+_0)-{26\over 24}{1\over\s}).\eqn\traj$$
For the rate of recession of the horizon we have (note that $\s^-$ ranges from
$0$ to $-\infty$)
$${d\s^-_H\over d\s^+}=-{a\over \l^2}\d (\s^+-\s^+_0)+{26\over
24}{1\over\l^2\s^{+2}}.\eqn\rate$$

The first term is the contribution of the infalling ($a>0$) matter which tends
to make the horizon move forward and the second (positive defintite) term is
that due to Hawking radiation and causes the horizon to recede. This equation
should be compared with equation (21) of [\lenny ]. It should be noted that the
 factor $N$ in the latter is replaced by 26 and $1\over\s^{+2}_0$ by
$1\over\s^{+2}$. (The in fall term is not written out in [\lenny ]).

Again following [\lenny ] we may define the mass of the black hole as
$M(\s^+ )=\l e^{-2\phi}|_H$, where the right hand side is to be evaluated at
the apparent horizon. For the classical solution this expression gives the ADM
mass of the black  hole. Using $\pa_-(e^{-2\phi})=-\l^2\s^+$ for large light
cone times, we have ${dM\over d\s^+}=-\l^3\s^+{d\s_H^-\over d\s^+}=-\l{26\over
24}{1\over\s^+}$. Or transforming to our asymptotically Minkowski coordinate
$\hat\s^+$,

$${dM\over d\hat\s^+}=-\l^2{26\over 24}.\eqn\decay$$

The rate of Hawking radiation can also be calculated directly by transforming
from the asymptotically Minkowski coordinate $\hat\s^-$ to the asymptotically
Minkowski coordinate $\ov\s=-{1\over\l}\log (e^{-\l\hat\s_-}-{a\over\l})$ in
which the classical solution is static. This coordinate covers only the region
outside the horizon and is the one appropriate to an outside observer at a
fixed distance from the black hole. Using \trans\ with $\s$ replaced by $\ov\s$
we get
$$t_{--}(\ov\s^-)=\l^2{26\over 24}\left [1-{1\over (1+ae^{\l\ov\s^-})^2}\right
]$$

in agreement at late light cone times with the rate of decay of mass.

 The important point is that  our analysis shows, that the
rate of decay of the apparent horizon, and the flux of Hawking radiation, are
governed by the ghost conformal
anomaly, but is of opposite sign (and hence the right sign for positive Hawking
radiation)
to what one might naively expect from  arguments which do not correctly take
back reaction into account. The problem with those arguments
can be seen from the second relation in \stress\ for the stress tensor. In the
usual semi-classical argument only the last term (proportional to $\k$) is
taken as radiation (caused by quantum anomalies). However when one substitutes
the exact solution into the "classical part"  there are terms $O(e^{2\phi})$
in the solution which when multiplied by the overall factor $e^{-2\phi}$ gives
terms which are of the same order as the term which is explicitly proportional
to $\k$. It is unclear which of these are to be considered back reaction terms
and which are radiation. By contrast our assumption of no ghosts in the
asymptotically Minkowski frame which covers the whole space is unambiguous.

One curious feature however is that (see \traj ) the apparent horizon goes over
asymptotically to the classical horizon as $\s^+\rightarrow 0$. Thus  a horizon
remains even though there is no singularity behind it (as we argued earlier).
One could also consider (primordial) black holes by taking $a=0=b_{\pm }$ in
\usoln, but putting $a_++a_-={M\over \l}$. In this case we see from \hor\ that
the horizon recedes all the way to end of space-time $\s^-=0$. One remaining
puzzle however is that the rate of Hawking radiation and the rate of decay of
mass are asymptotically constant (in the $\ov\s$ frame).It is not clear why
the radiation  stops (as it should when the mass goes to zero). Perhaps this is
related to our choice
of asymptotically Minkowski coordinates which was governed by the classical
solution. This probably needs to be corrected but at this point it is unclear
how to do this.

Finally we note that the physics is essentially determined by the Liouville
like
theory \newaction. In particular none of our conclusions would be changed if we
had used the proposal of Strominger [\andy ] which would have slightly modified
the relation between $X,Y$ and $\phi,\r$ (see [\sda ], [\bilal ]).

{\bf Acknowledgements}: I wish to thank Joe Polchinski for a discussion. This
work is partially
supported by Department of Energy contract No.
DE-FG02-91-ER-40672.

{\bf Note added.} The Bondi (or ADM) mass of our solutions can be computed
directly from the expressions for the stress tensors  in terms of the fields
$X,Y$.

  $$\eqalign{T_{\pm\pm}=&-{1\over
2}(\pa_{\pm}X\pa_{\pm}X-\pa_{\pm}Y\pa_{\pm}Y)-\sqrt{\k\over 2}\pa^2_{\pm}Y
 +{1\over 2}\sum_i\pa_{\pm}f^i\pa_{\pm}f^i\cr
T_{+-}=&\sqrt{\k\over
2}\pa_+\pa_-Y-\l^2e^{-\sqrt{2\over\k}(X-Y)}.\cr}\eqn\enmom$$

To find the mass we need to linearize around a reference static solution in
asymptotically Minkowski coordinates. In our original  coordinate system in
which $g_{\pm}$ are zero we may choose such a static solution to be
$$X=Y=\sqrt{2\over\k}(\l^2\s^+\s^-+{26\over 24}\log (-\s^+\s^-)),~ f=0.$$.

This is obtained from equations (13) and (24) by putting $a_{\pm}=b_{\pm}=a=0$
in (24). Transforming to the asymptotically Minkowski coordinates $\ov\s^{\pm}$
defined by $\s^+={1\over\l}e^{\l\ov\s^+}, \s^-=-{1\over\l}e^{-\l\ov\s^-}$,
$$\eqalign{X=&-\sqrt{2\over\k}\left (e^{\l(\ov\s^+-\ov\s^-)}-{26\over
24}\l(\ov\s^+-\ov\s^-)+{26\over 24}\log\l^2\right )\cr
=&Y-\sqrt{\k\over 2}\l (\ov\s^+-\ov\s^-).\cr}\eqn\static$$
This solution corresponds to the classical linear dilaton solution. Linearizing
around this
solution we have the following expression for the Bondi
 mass

$$\eqalign{M(\ov\s^-)=&-\int d\ov\s^+(\d T_{++}+\d T_{+-})\cr
=&-\sqrt{2\over \k}\l e^{(\ov\s^+-\ov\s^-)}(\d X-\d Y)+
{26\over 24}\l (\d X-\d Y)\cr &-\sqrt{\k\over 2}\l\d Y+
\sqrt{\k\over 2}(\pa_+\d Y-\pa_-\d Y).\cr}\eqn\bondi$$

It is easily seen that this expression (apart from the ghost terms), goes over
to equation (26) of [1] in the semi-classical limit $\k e^{2\phi}<<1$.

First we can look for static solutions corresponding to black holes. In this
case  we must put $a=b_{\pm}=0$ and $a_{\pm}\ne 0$. Then for $\ov\s^+>>1$,
$\d Y=\sqrt{2\over \k}(a_++a_-)=\d X$ and we have from \bondi,
$$M(\ov\s^-)= \l (a_++a_-)$$
So this solution corresponds to black holes with constant mass.
Now let us consider the collapsing matter solutions with $a_{\pm}=b_{\pm}=0,
a\ne 0$. In this case
$$\d X=\d Y=-\sqrt{2\over\k}\left [ae^{\l\ov\s^+_0}-\t
(\ov\s^+-\ov\s^+_0)-{26\over 24}\log
\left (1+{a\over\l}e^{+\l\ov\s^-}\t (\ov\s^+-\ov\s^+_0)\right )\right ]$$

Substituting in \bondi\ and taking the derivative with respect to $\ov\s^-$,
we have for the rate of decay of  Bondi mass,

$${dM\over d\ov\s^-}\biggr\vert_{\ov\s^-\rightarrow\infty}=-{26\over 24}\l^2$$
 in agreement with the rate of Hawking radiation calculated earlier.
 The expression for the the mass itself is,
 $$M(\ov\s^- )=a\l e^{\l\ov\s^+_-}-{26\over 24}\l\log
(1+{a\over\l}e^{\l\ov\s^-})-{26\over
24}{\l\over 1+{\l\over a}e^{-\l\ov\s^-}}.$$
  This expression resolves the puzzle that we mentioned in the discussion after
equation in the penultimate paragraph of the paper. The Hawking radiation does
not stop because the Bondi mass goes all the way to negative infinity! This is
not
surprising since as we saw from the static solutions there are solutions with
arbitrarily negative mass ( $a_++a_-$ can be chosen to be negative). This is
not an artifact of the quantization of the
  theory. Even the classical dilaton gravity theory has negative mass solutions
  and the theory has no ground  state. It should be noted that there is no
positive energy theorem for dilaton gravity in 2 dimensions since the dilaton
  kinetic energy has the wrong sign. (In higher dimensions it is possible to
perform a Weyl transformation to the canonical metric in terms of which the
dilaton kinetic energy has the right sign). Of course in the classical theory
negative mass solutions have naked singularities whereas as we have shown the
quantum corrections tame these singularities.

  Even though the theory has no ground state it nevertheless illustrates
explicitly several important effects that one expects a more realistic theory
to satisfy,
  namely the taming of classical singularities by quantum effects, and the back
reaction of the Hawking radiation which causes the mass of the black hole to
decay.

Acknowledgement: I wish to thank Andy Strominger for an e-mail message

on work done by him and Steve Giddings.

\refout
\end